\documentclass[%
reprint,
 amsmath,amssymb,
 aps,
 pre,
]{revtex4-1}

\usepackage{graphicx}
\usepackage{dcolumn}
\usepackage{bm}
\usepackage{float}
\usepackage{color}
\usepackage[percent]{overpic}

\def\figlen/{320} 

\begin{document}
\title{Characterization of phase transitions in a model ecosystem of sessile species}
\author{Florian Uekermann}
\author{Joachim Mathiesen}
\author{Namiko Mitarai}
\email{mitarai@nbi.dk}
\affiliation{Niels Bohr Institute, Blegdamsvej 17, 2100 Copenhagen, Copenhagen University, Denmark}

\date{\today}

\begin{abstract}
We consider a model ecosystem of sessile species competing for space. In particular, we consider the system introduced in [Mathiesen et al. Phys. Rev. Lett. 107, 188101 (2011)] where species compete according to a fixed interaction network with links determined by a Bernoulli process. In the limit of a small introduction rate of new species, the model exhibits a discontinuous transition from a high-diversity state to a low-diversity state as the interaction probability between species, $\gamma$, is increased from zero. Here we explore the effects of finite introduction rates and system-size on the phase transition by utilizing efficient parallel computing.
We find that the low state appears for  $\gamma>\gamma_c$. As $\gamma$ is increased further, the high state approaches to the low state, suggesting the possibility that the two states merge at a high $\gamma$. We find that the fraction of time spent in the high state becomes longer with higher introduction rates, but the availability of the two states is rather insensitive to the value of the introduction rate. Furthermore we establish a relation between the introduction rate and the system size, which preserves the probability for the system to remain in the high-diversity state.
\end{abstract}

\pacs{Valid PACS appear here}
\maketitle

\section{Introduction}
Many ecosystems maintain a high diversity in spite of the competition among species, {and mechanisms of coexistence remain a central question in ecology. Especially, the seminal works by Gardner and Ashby \cite{gardner1970connectance} and May \cite{may-1972} that pointed out in a simple toy model that a well-mixed system with random interaction inhibits stable coexistence made researchers explore various explanations for the development of stable ecosystems \cite{roberts1974stability,pimm1984complexity,mccann2000diversity,ives}}. For example, a non-hierarchical interaction or cyclic relationship among species has been proposed as a way to obtain long-lasting coexistence \cite{itoh1971boltzmann,may1975nonlinear,hofbauer1998evolutionary,reichenbach2006coexistence,claussen2008cyclic}, though it often causes oscillatory dynamics of the population. When space is explicitly considered, the oscillation of the total population can be suppressed and instead give rise to a spiral wave-like dynamics, which further stabilizes the coexistence of species \cite{dayton,gilpin,jackson-1975,karlson,tainaka1988lattice,boerlijst-1991,cronhjort,Szabo2001,kerr-2002,murrell-2003,kirkup2004antibiotic,Laird2006,kerr,reichenbach2008instability,rulands2013global,szolnoki2016zealots}. 

In \cite{mathiesen2011ecosystems} a {simple spatial} ecosystem model was introduced for mutually exclusive and sessile species that compete for available space. The model was inspired by the coexistence of lichen species \cite{harris1996competitive,N08,G09,JNHTM10} and corals \cite{jackson-1975,AS1997,buss,karlson}. 
The model introduced a {competitive} interaction between occupants of neighboring sites on a two-dimensional square lattice.
A species can replace the occupant of a neighboring site if a predefined interaction network contains a link from that species to the species in the neighboring site. In the model, the interactions among species are assigned randomly, according to a Bernoulli process. When the connectivity of the interaction network increases, a transition from multiple coexisting species to one species is observed \cite{mathiesen2011ecosystems}.  For this model it was shown that the presence of cyclic interactions among species is necessary to maintain a high-diversity state  \cite{mitarai2012emergence}. In particular, cyclic interactions lead to a complex spatiotemporal dynamics that create spatial patches, which provide niches for a newly introduced species. In the limit of small introduction rates, the transition between the high- and low-diversity state was found to be sharp and discontinuous. For small, but finite, introduction rates the model shows a bistable dynamics, switching between low- and high-diversity states above the critical connectivity \cite{mitarai2012emergence}.
This behavior is found to be robust to changes in the lattice structure \cite{mathiesen2011ecosystems}, the introduction of a very small death rate \cite{botta2014disturbance} and to making the introduced new species mutations of existing ones \cite{mitarai2014speciation}. 
For larger introduction rates, the diversity tends to be higher and the bistability is smeared out by inherent noise in the dynamics. 

{ This is one of the simplest models of spatial coexistence of competitive species. The interaction network is completely random, and the model is characterized by the only two parameters, i.e., the connectivity of the interaction network and the introduction rate of new species. Understanding the properties of the transition between low- and high-diversity state will improve our basic understanding of the coexistence mechanism.
However,} previously, the characteristics of the transition were not carefully analyzed. In particular, it was unclear whether separate high- and low-diversity states exist for large introduction rates, or if they merge at a critical point. 

Here we examine the effect of the species introduction rate on the transition between low and high diversity. We first demonstrate that increasing the introduction rate gradually reduces the time spent in the low-diversity state, while the bistable dynamics is maintained. However, independent of the introduction rate, the model maintains a qualitative transition at a critical probability of interactions, below which the low diversity state is inaccessible.
The distinction between the high and the low diversity state tends to be smaller with higher interaction probability, again independent of the introduction rate.
Thus the introduction rate and frequency of interaction play orthogonal roles in how they affect the qualitative dynamics.
The accessibility of the low-diversity state is determined by the frequency of interaction, while the introduction rate controls the time spent in the respective state if the low-diversity state is accessible.
Finally, we show that scaling the introduction rate inversely proportional to the lattice length, the fraction of time spent in high and low state is conserved for different system sizes.

\begin{figure*}
\begin{overpic}[width=0.49\linewidth]{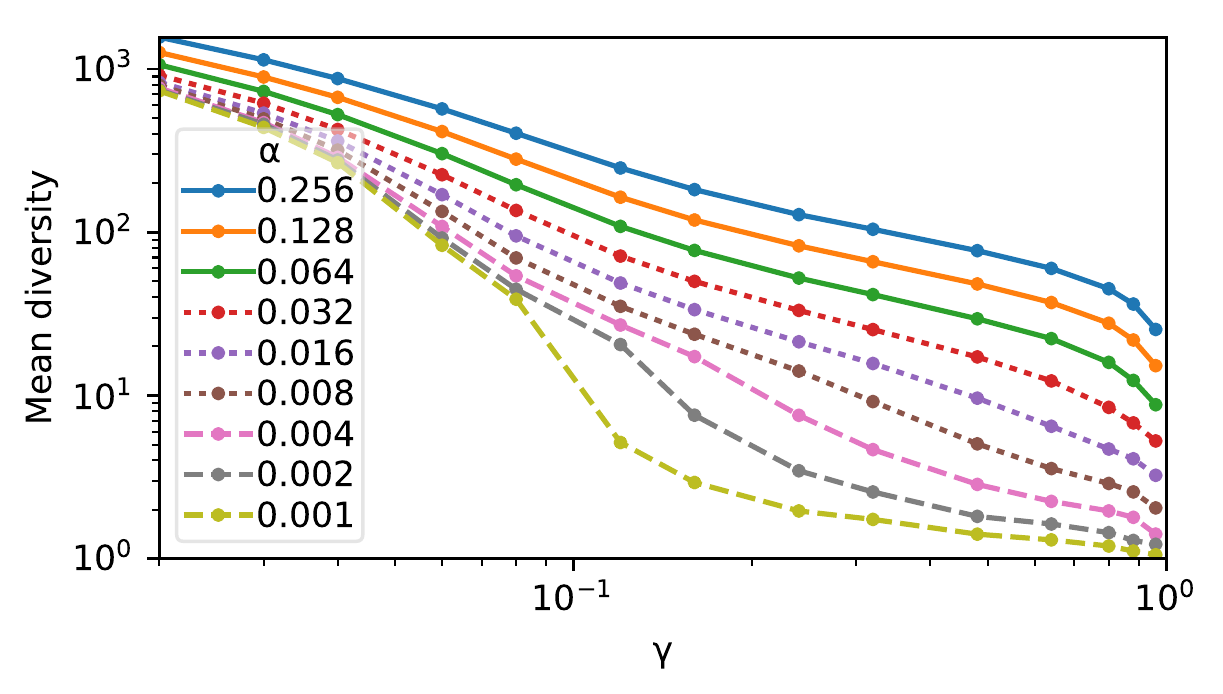}
\put (0,55) {\bf(a)}
\end{overpic}
\begin{overpic}[width=0.49\linewidth]{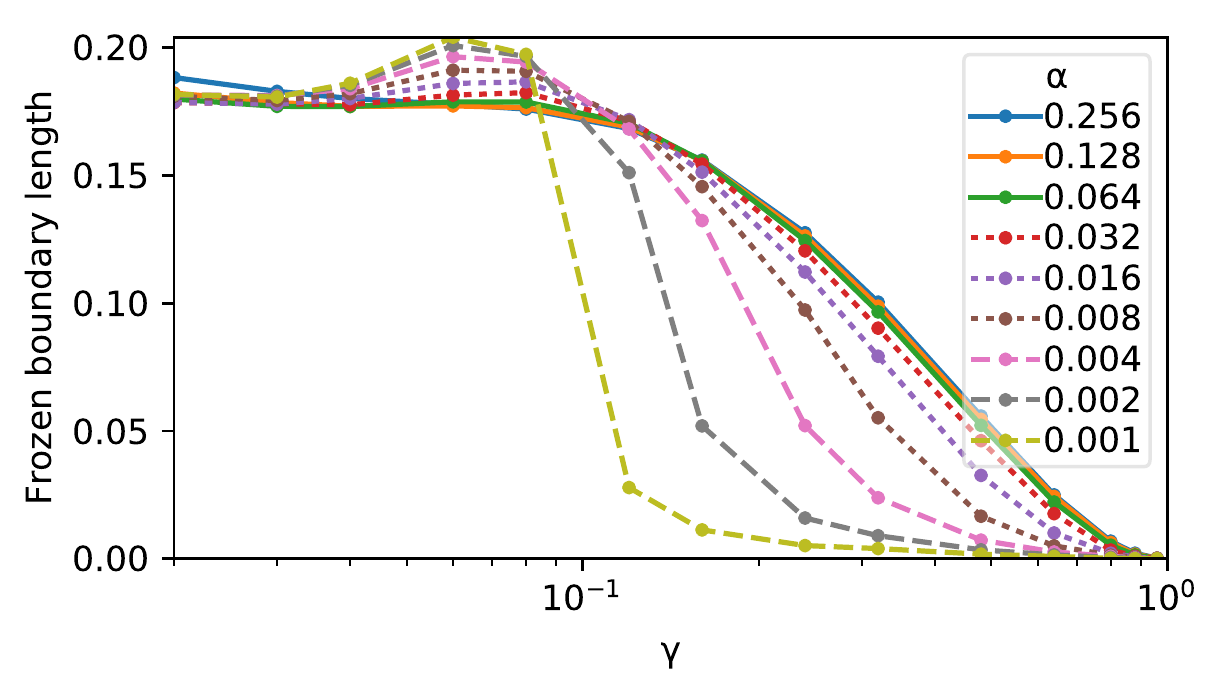}
\put (0,55) {\bf(b)}
\end{overpic}
\caption{\label{fig:classes}
Mean diversity (a) and mean length of frozen boundary (b) is shown as function of interaction frequency $\gamma$ for different introduction rates $\alpha$. $N=320^2$.}
\end{figure*}

The paper is structured as follows:
In Sec. \ref{sec:model}, we briefly describe the model\cite{mitarai2012emergence,mathiesen2011ecosystems} and the algorithm for parallel computing. We also give precise definitions of the terms introduction rate and interaction frequency. Sec. \ref{sec:boundary} introduces the frozen boundary length as a more robust classification criterion for low and high states, compared to the diversity.
In Secs. \ref{sec:high-low} and \ref{sec:fraction}, we examine how the interaction frequency, introduction rate and system size influence the availability and prevalence of the low state. Finally, we observe a continuous behavior at the phase-transition for finite introduction rates.

\section{Model\label{sec:model}}
In this simple ecosystem model, multiple species compete for occupation of space on a 2-dimensional square-lattice.
A lattice site can only be occupied by one species at a time.
The interactions in the ecosystem are defined by a directed network of {competitive} interactions between species.
A species can invade a neighboring site and replace its occupant if the interaction network contains a link from the invading species to the occupying species.
The interaction network is described by $\Gamma(s_i,s_o) \in \{0,1\}$, where $s_i$ and $s_o$ represent the invading species and occupying species, respectively.
$\Gamma(s_i,s_o)=1$ describes a link from species $s_i$ to $s_o$, which means $s_i$ can replace $s_o$ on a neighboring lattice site.
Each element of the matrix $\Gamma$ is predetermined to be $1$ with probability $\gamma$, or $0$ with probability $1-\gamma$. In this model, there are two ways for species $s_i$ and $s_j$ to be competitively equivalent: One way is $\Gamma(s_i,s_j)=\Gamma(s_j,s_i)=1$, i.e. the species can replace each other, the other is $\Gamma(s_i,s_j)=\Gamma(s_j,s_i)=0$, i.e. they cannot replace each other (stand-off). 
{ Note that both cases are observed in the nature; 
mutual active competition was reported in the epifaunal communities \cite{KayandKeough1981}, while  for crustose lichens, a contact boundary is formed at the encounter of one with another, and if they are competitively neutral these boundaries may remain stable over time \cite{JNHTM10}.}

Previous implementations \cite{mitarai2012emergence,mathiesen2011ecosystems} used a random sequential update scheme, consisting of repeatedly picking two neighboring sites as source and target of a potential invasion.
To utilize the highly parallel architecture of modern graphics processing units, we shall here use a modified updating scheme.
The square lattice is divided into two sets of $2\times2$-chunks.
Each set consists of all chunks that are not directly adjacent to each other, resembling the set of black and white sites of a checkerboard.
In each chunk of one set a random site is picked as target for the potential invasion and a random neighbor is picked as source.
If the interaction network contains a link from the invading species to the one occupying the target site, the invading species becomes the occupant of the target site.
This update step is repeated alternatingly for the two sets of chunks (black and white).
This update mechanism was chosen to allow parallel updates that guarantee that the origin of an invasion is never invaded at the same time (GPU friendly), while preserving a local dynamics similar to random sequential updates. 
We have found no qualitative or quantitative differences between these simulations and those with random sequential updates.

The rate of introducing new species is parametrized by the introduction rate $\alpha$.
A new species is introduced at a random site with a rate of $\alpha L^{-2}$, where $L$ represents the length of the lattice.
For each potential interaction between the new and existing species a link is added with probability $\gamma$.
Additionally the interaction network will always contain a link from the new species to the former occupant of the site at which it is introduced.
This definition of $\alpha$ differs from our previous papers \cite{mitarai2012emergence,mathiesen2011ecosystems}, were we chose $\alpha$ to be the rate of attempted introductions of a new species, which were rejected if the interaction network did not have a link from the new species to the old one.
Thus, the $\alpha$ value used in this paper would have to be divided by $\gamma$ when one compares it with previous results.

\begin{figure}
\begin{overpic}[width=\linewidth]{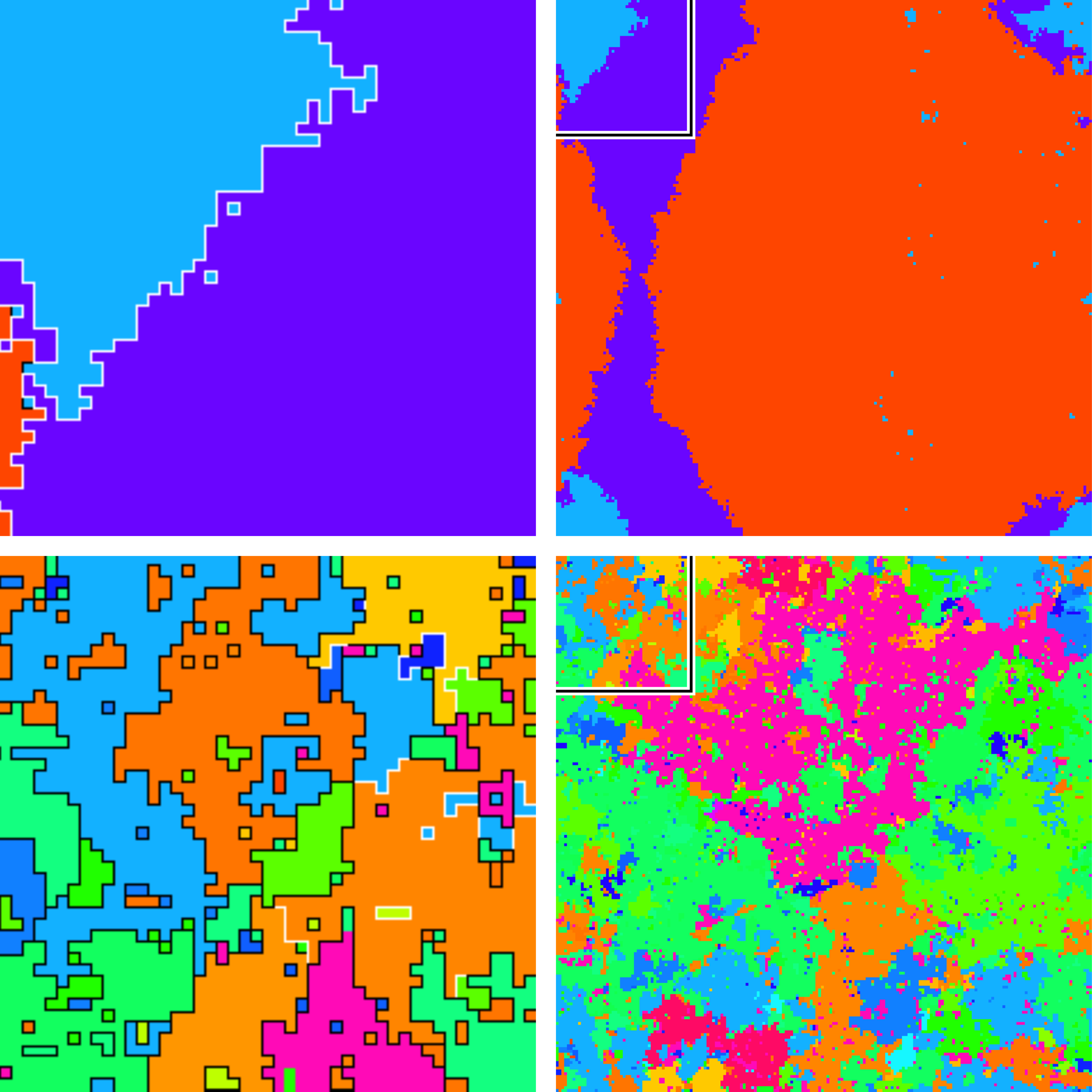}
\put (7,97) {\bf\color{white}(a)}
\put (89,97) {\bf\color{white}(b)}
\put (7,46) {\bf\color{white}(c)}
\put (89,46) {\bf\color{white}(d)}
\end{overpic}
\caption{\label{fig:boundary}
Snapshots of system in low (a,b) and high (c,d) state. (a) and (c) show magnification of upper left corners of the snapshots (b) and (d), respectively, with marked frozen (black) and active (white) boundaries. $N=192^2$, $\gamma=0.7$, $\alpha=0.0005$.
}
\end{figure}
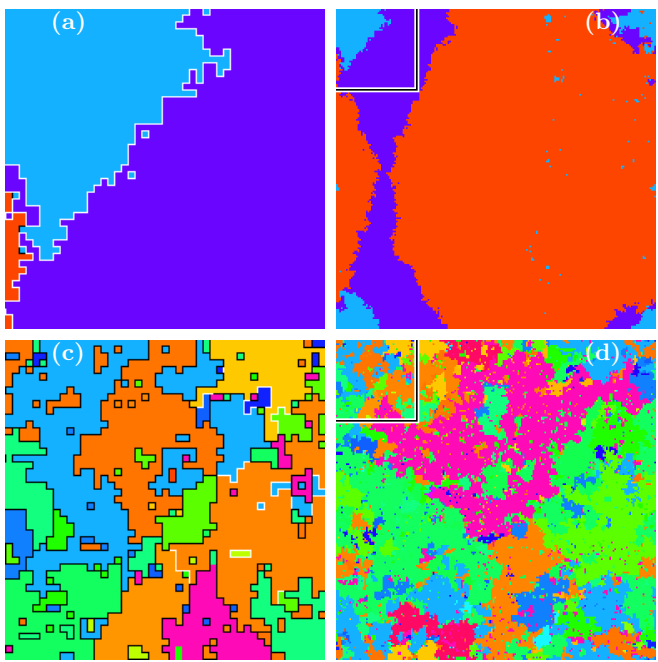

\section{Results}

\subsection{Diversity and frozen boundary length\label{sec:boundary}}

\begin{figure}
\includegraphics[width=\linewidth]{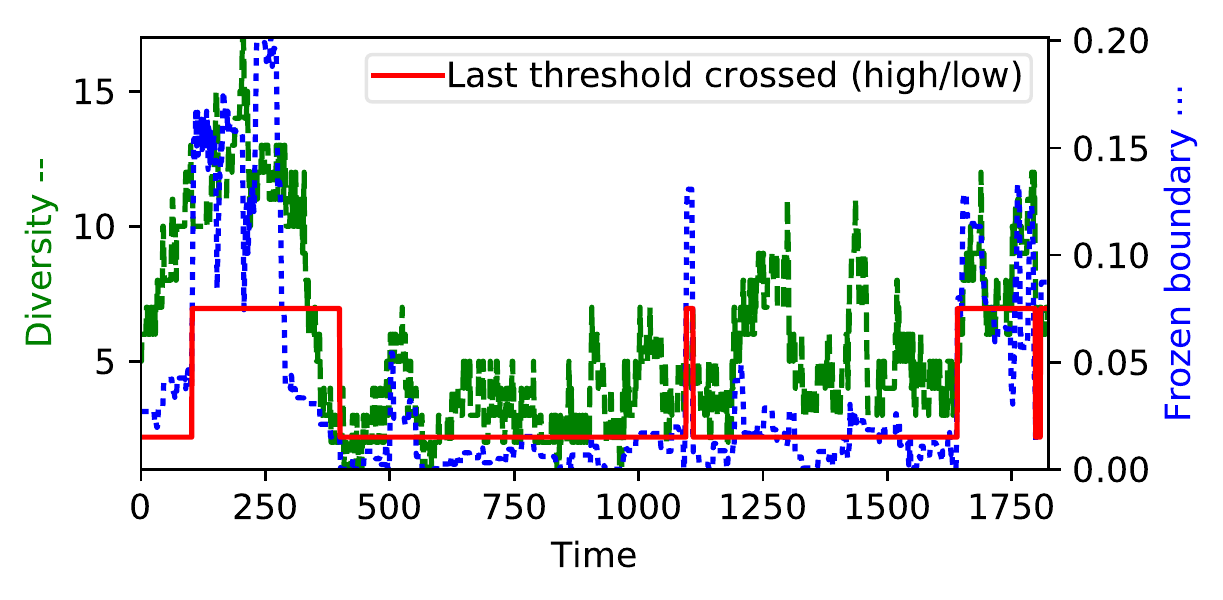}
\caption{\label{fig:classifier}
Time series of diversity and frozen boundary length with introduction rate $\alpha=0.003$ and interaction frequency $\gamma=0.16$ in a $N=192^2$ system. For higher $\gamma$ and $\alpha$ the diversity (green, dashed) fluctuations are stronger and the frozen boundary length (blue, dotted) is a more reliable criterion. The solid red curve shows whether the system is classified to be in high or low state, based on the frozen boundary length. The curve shows frozen boundary length threshold value that was crossed for the current classification.
}
\end{figure}

In previous work the diversity of species, {defined as the number of species}, was used as the main observable to quantify the state of the system. The mean diversity shows a sharp transition 
from high- to low- values at $\gamma\approx0.055$ when $\alpha$ is close to zero and $N=192^2$ \cite{mitarai2012emergence,mathiesen2011ecosystems} as shown in Figure~\ref{fig:classes}(a).  However, the fluctuation of the diversity turned out to be significant for larger $\alpha$ (see Fig.~\ref{fig:classifier}(a), green dashed line), making the classification of the two states difficult.  

It was found that the transition to high or low diversity state is preceded by an increase or decline in the number of patches (spatially connected sites occupied by the same species) \cite{mitarai2012emergence}, which can also be used as an indicator of the high- and low-state. Among those patches, the most important ones for high-diversity are the patches surrounded by the species with stand-off relations, since those patches will be long-lasting.  This can be clearly seen in Fig.~\ref{fig:boundary} where the active boundaries between interacting species and the frozen boundaries between stand-off relation species are marked in the snapshots of the high- and low-state.
We found that the length of boundaries between sites with different species of stand-off relationships (Fig.~\ref{fig:classifier}(a), blue dotted line) a is better indicator of the high- and low-state with less fluctuation than the diversity (Fig.~\ref{fig:classifier}(a), green dashed line), and shows the transition between the high- and the low- state at the same position as the mean diversity for small $\alpha$ (Fig.~\ref{fig:classes}(b)). 
We therefore use this quantity, normalized with its maximum possible value ($2L^2$), to study the transition in finite $\alpha$, and call it the frozen boundary length $F_l$.

\subsection{High-state and Low-state \label{sec:high-low}}

\begin{figure}
\begin{overpic}[width=\linewidth]{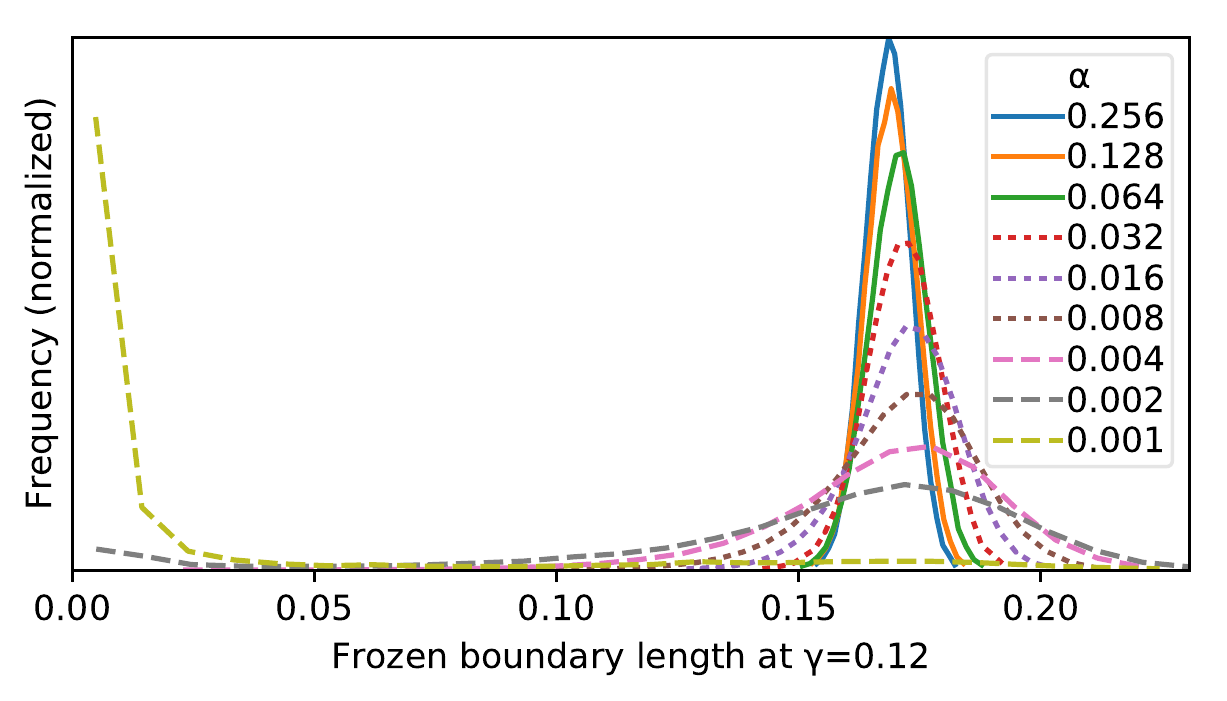}
\put (0,55) {\bf(a)}
\end{overpic}
\begin{overpic}[width=\linewidth]{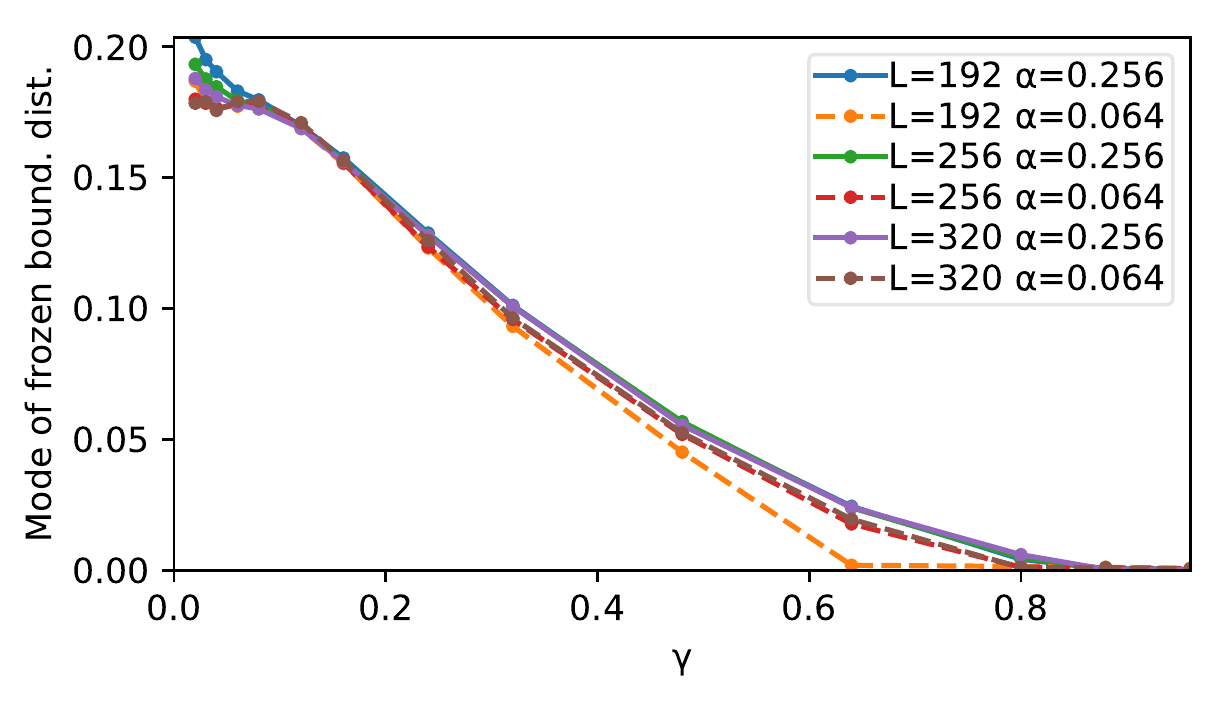}
\put (0,55) {\bf(b)}
\end{overpic}
\caption{\label{fig:frzDist} (a) Distribution of boundary lengths with $N=320^2$ $\gamma=0.12$ for various $\alpha$. (b) Location of maximum (mode) as a function of $\gamma$.}
\end{figure}

Figure~\ref{fig:frzDist}(a) shows the distribution of the frozen boundary length $F_l$ for different introduction rates $\alpha$ at $\gamma=0.12$.
With growing $\alpha$, the distribution gains a high mode.
For lower values of $\alpha$, the peak at large $F_l$ lowers its height, and eventually becomes undetectable.
For high $\alpha$, the peak at low $F_l$ peak becomes undetectable.
For intermediate values (dashed grey line), the distribution is bimodal.
This is consistent with the observation in previous work, that in the limit of $\alpha\to 0$ no bi-stability is observed and the system is exclusively in low state when $\gamma$ exceeds the critical value $\gamma_c$.

For a fixed $\alpha$, the peak position of the high mode decreases to $F_l\approx 0$ with increasing $\gamma$
as shown in Fig.~\ref{fig:frzDist}(b), while the left peak in low $F_l$ stays always around $F_l=0$ if it is there. 
Eventually, for high enough $\gamma$ the two peaks becomes indistinguishable within our numerical accuracy.
Figure~\ref{fig:frzDist}(b) also shows that the high peak's position is rather insensitive to the system size $L$ and the value of $\alpha$. 

In the following we will use the location of the peak of the right mode of the frozen boundary length $m$ of a given $\gamma$ and $L$ with $\alpha=0.256$ to classify whether the system is in high and low state. More specifically, we record the time series of the frozen boundary length $F_l$ (e.g., Fig.~\ref{fig:classifier}(a), blue dotted line), and if $F_l$ is longer than $0.5m$, then the system is classified to have entered a high state samples, until the length of the boundary drops below $0.1m$ and vice versa. This allows us to classify the state of the system, as shown by the red curve in Fig.~\ref{fig:classifier}(a).

\subsection{Fraction of time in high- and low-state\label{sec:fraction}}
\begin{figure*}[bth]
\begin{overpic}[width=0.49\linewidth]{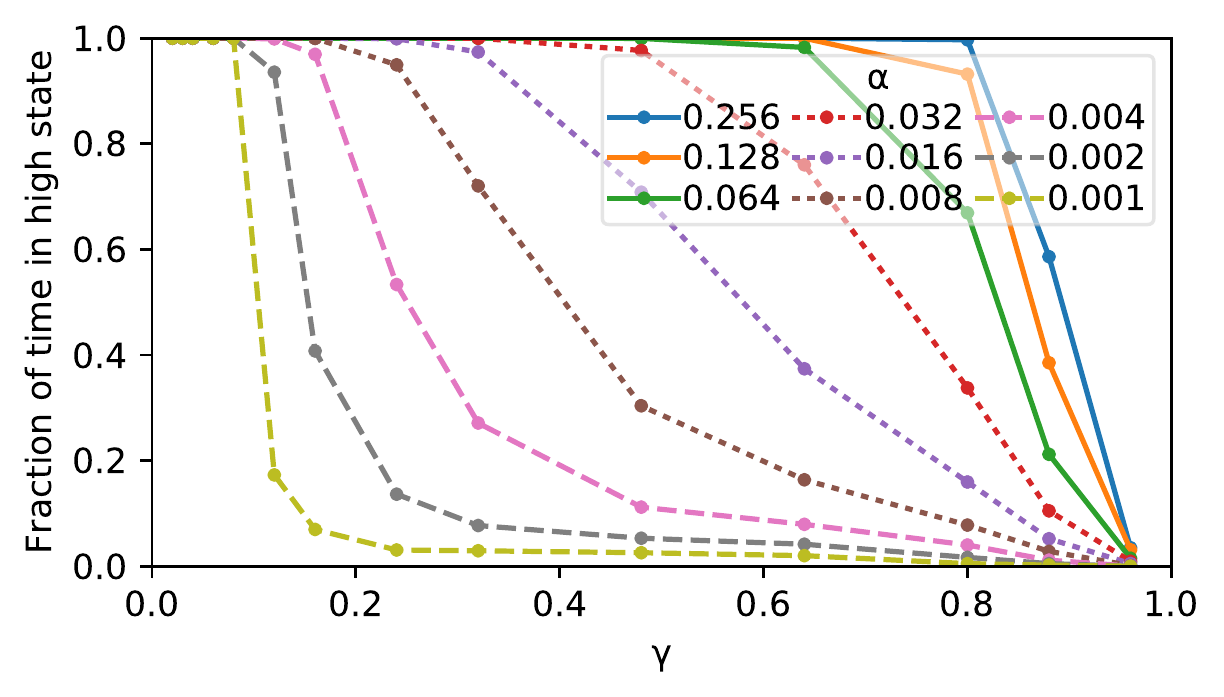}
\put (0,55) {\bf(a)}
\end{overpic}
\begin{overpic}[width=0.49\linewidth]{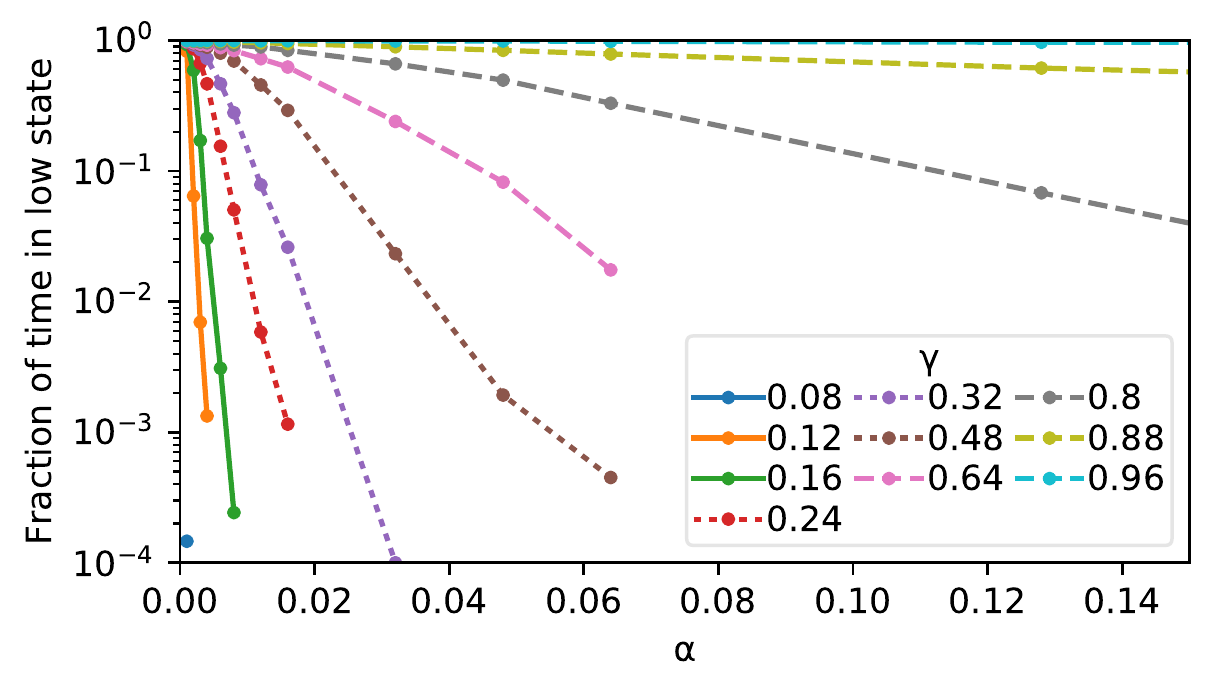}
\put (0,55) {\bf(b)}
\end{overpic}
\caption{\label{fig:p} Fraction of high (a) and low (b) state samples over $\gamma$ and $\alpha$. $N=320^2$.}
\end{figure*}

Using the frozen boundary length as classification criterion as described in the previous section, we classified the samples of simulations with different choices for $L\in\{192,256,320\}$ and various $\alpha$ and $\gamma$.
Fig. \ref{fig:p} shows our results for $L=320$.

As observed previously \cite{mitarai2012emergence}, the clearly separated low state becomes available at $\gamma>\gamma_c$ when $\alpha$ is small enough. The critical value  was found to be approximately  $\gamma_c\approx0.055$, for $L=192$ \cite{mitarai2012emergence,mathiesen2011ecosystems}. For larger systems $L\in{256, 320}$ we find larger critical values $\gamma_c\approx0.1$.

With increasing $\gamma$ beyond $\gamma_c$, the fraction of time spent in high state approaches $0$ in a continuous fashion, and the fraction increases with $\alpha$ as shown Fig. \ref{fig:p}(a). We observe that the fraction of time spend in high-state approaches a step-like shape as a function of $\gamma$ with a sharp change at $\gamma_c$ as the limit $\alpha\to 0$ is approached, which is consistent with the results of quasistatic simulations \cite{mathiesen2011ecosystems} that showed a discontinuous transition.

When plotting the time spent in low state over $\alpha$ for fixed $\gamma$ (fig. \ref{fig:p}(b)),
we find that with increasing $\alpha$ the time in low state decreases faster than exponentially at first, but seems to approach an exponential decline as $\alpha$ increases.
This behavior is observed for all $\gamma>\gamma_c$. At lower $\gamma$ we did not find any low state samples regardless of $\alpha$.

In practice we can measure this decline of time spent in low state only over few orders of magnitude ($2$ to $4$), due to limitations on simulation time.
Consequently, the range of $\alpha$ values where we can measure this exponential decline is limited.
This applies especially at low $\gamma$ values, that exhibit quicker decline (larger absolute exponent).
However, our observations do not indicate any limit to the exponential decline with increasing $\alpha$ values.
This suggests that the $\gamma$ value governs whether the low state is available at all ($\gamma>\gamma_c$), while $\alpha$ is an orthogonal parameter, that determines how much time the system spends in low state if $\gamma>\gamma_c$.

\begin{figure}
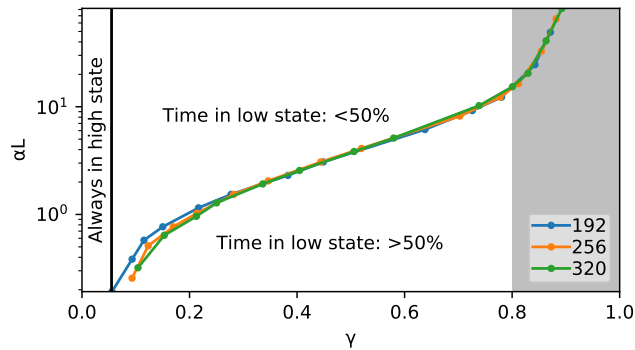

\includegraphics[width=\linewidth]{{{trans}.pdf}}
\caption{\label{fig:phasediag} Phase diagram. Classification threshold for the frozen boundary is questionable in grey area due to boundary length approaching zero.}
\end{figure}

For different system sizes we observe the same qualitative behavior (we tested $L=192$, $L=256$ and $L=320$).
However, we observe that $\alpha$ needs to be scaled inversely proportional to the system size to achieve a quantitatively similar amount of time spent in low and high state (see Fig. \ref{fig:phasediag}).

The scaling of $\alpha$ with $L$ can be understood as follows: In a low-state, typically there is only one species in a system. When a new species is introduced, it will take time proportional to $L$ for a new species to replace the old one. If the next new species is introduced after this displacement, there is no chance for the system to go high-state. In other words, $\alpha L$ determines the rate for the system state to switch from low to high.

\section{Discussion}
Through extensive numerical simulations, we have considered the transition between high and low diversity states in a model ecosystem. The simulations show that there is a clear discontinuous transition at $\gamma=\gamma_c$, above which the low state, clearly separated from the high state, becomes available independent of the value of $\alpha$. In the $\alpha\to 0$ limit, the system stays in the low state for $\gamma>\gamma_c$, but with finite $\alpha$,  we have confirmed the bistability between the high state and the low state. 

Interestingly, the high state seems to approach the low state as $\gamma$ approaches to 1, rather independently from $\alpha$. It is possible that the high state appears continuously from the low state at a certain $\gamma$.  When $\gamma=1$, i.e., when every species can invade everyone else, the frozen boundary length must be zero and there should not be any distinction between the high state and the low state. It would be interesting to confirm at which value of $\gamma$ the high state appears, when decreasing $\gamma$ from 1; it may continuously appear at $\gamma=1$, or there there may be another transition to the high state at $\gamma<1$. In the latter case, the high state may appear continuously from the low state, discontinuously with small separation between the two states. This is difficult to confirm numerically, because of fluctuations and the small gap between the low and the high states at high $\gamma$ we see in our simulations. It should also be noted that our result cannot exclude some dependency on $\alpha$ of the high state at high $\gamma$, since identification of the high peak is less accurate at high $\gamma$'s. Further investigation of the nature of this transition likely depends on the development of analytical insight. 

{ In the present work, we could not identify a critical point where the two states merges, which would have been expected in a more conventional phase transition. Further research is needed to clarify the nature of the transition in the present model.}

Overall, our result suggests that the available states are solely controlled by $\gamma$, while $\alpha$ controls the fraction of time spent in each state. This may suggest that the available states are determined by geometrical constraint such as how many neighbor species exist and can interact for a given $\gamma$. It would be interesting to see the transition in three dimensions where the geometrical constraints is entirely different from those in two dimensions or on a different lattice. { It could provide us better understanding of the mathematical structure behind the transition. It may also provide some insight about coexistence of microbial community, which is often studied in two dimensional lattice models with various invasion rules \cite{czaran2002chemical,kerr2002} but can form three dimensional structures in e.g., soil and biofilms.}

\section*{Acknowledgement}
This work is supported by Danish National Research Foundation. 

\bibliography{bibliography}

\begin{thebibliography}{40}%
\makeatletter
\providecommand \@ifxundefined [1]{%
 \@ifx{#1\undefined}
}%
\providecommand \@ifnum [1]{%
 \ifnum #1\expandafter \@firstoftwo
 \else \expandafter \@secondoftwo
 \fi
}%
\providecommand \@ifx [1]{%
 \ifx #1\expandafter \@firstoftwo
 \else \expandafter \@secondoftwo
 \fi
}%
\providecommand \natexlab [1]{#1}%
\providecommand \enquote  [1]{``#1''}%
\providecommand \bibnamefont  [1]{#1}%
\providecommand \bibfnamefont [1]{#1}%
\providecommand \citenamefont [1]{#1}%
\providecommand \href@noop [0]{\@secondoftwo}%
\providecommand \href [0]{\begingroup \@sanitize@url \@href}%
\providecommand \@href[1]{\@@startlink{#1}\@@href}%
\providecommand \@@href[1]{\endgroup#1\@@endlink}%
\providecommand \@sanitize@url [0]{\catcode `\\12\catcode `\$12\catcode
  `\&12\catcode `\#12\catcode `\^12\catcode `\_12\catcode `\%12\relax}%
\providecommand \@@startlink[1]{}%
\providecommand \@@endlink[0]{}%
\providecommand \url  [0]{\begingroup\@sanitize@url \@url }%
\providecommand \@url [1]{\endgroup\@href {#1}{\urlprefix }}%
\providecommand \urlprefix  [0]{URL }%
\providecommand \Eprint [0]{\href }%
\providecommand \doibase [0]{http://dx.doi.org/}%
\providecommand \selectlanguage [0]{\@gobble}%
\providecommand \bibinfo  [0]{\@secondoftwo}%
\providecommand \bibfield  [0]{\@secondoftwo}%
\providecommand \translation [1]{[#1]}%
\providecommand \BibitemOpen [0]{}%
\providecommand \bibitemStop [0]{}%
\providecommand \bibitemNoStop [0]{.\EOS\space}%
\providecommand \EOS [0]{\spacefactor3000\relax}%
\providecommand \BibitemShut  [1]{\csname bibitem#1\endcsname}%
\let\auto@bib@innerbib\@empty
\bibitem [{\citenamefont {Gardner}\ and\ \citenamefont
  {Ashby}(1970)}]{gardner1970connectance}%
  \BibitemOpen
  \bibfield  {author} {\bibinfo {author} {\bibfnamefont {M.~R.}\ \bibnamefont
  {Gardner}}\ and\ \bibinfo {author} {\bibfnamefont {W.~R.}\ \bibnamefont
  {Ashby}},\ }\href@noop {} {\bibfield  {journal} {\bibinfo  {journal}
  {Nature}\ }\textbf {\bibinfo {volume} {228}},\ \bibinfo {pages} {784}
  (\bibinfo {year} {1970})}\BibitemShut {NoStop}%
\bibitem [{\citenamefont {May}(1972)}]{may-1972}%
  \BibitemOpen
  \bibfield  {author} {\bibinfo {author} {\bibfnamefont {R.~M.}\ \bibnamefont
  {May}},\ }\href@noop {} {\bibfield  {journal} {\bibinfo  {journal} {Nature}\
  }\textbf {\bibinfo {volume} {238}},\ \bibinfo {pages} {413} (\bibinfo {year}
  {1972})}\BibitemShut {NoStop}%
\bibitem [{\citenamefont {Roberts}(1974)}]{roberts1974stability}%
  \BibitemOpen
  \bibfield  {author} {\bibinfo {author} {\bibfnamefont {A.}~\bibnamefont
  {Roberts}},\ }\href@noop {} {\bibfield  {journal} {\bibinfo  {journal}
  {Nature}\ }\textbf {\bibinfo {volume} {251}},\ \bibinfo {pages} {607}
  (\bibinfo {year} {1974})}\BibitemShut {NoStop}%
\bibitem [{\citenamefont {Pimm}(1984)}]{pimm1984complexity}%
  \BibitemOpen
  \bibfield  {author} {\bibinfo {author} {\bibfnamefont {S.~L.}\ \bibnamefont
  {Pimm}},\ }\href@noop {} {\bibfield  {journal} {\bibinfo  {journal} {Nature}\
  }\textbf {\bibinfo {volume} {307}},\ \bibinfo {pages} {321} (\bibinfo {year}
  {1984})}\BibitemShut {NoStop}%
\bibitem [{\citenamefont {McCann}(2000)}]{mccann2000diversity}%
  \BibitemOpen
  \bibfield  {author} {\bibinfo {author} {\bibfnamefont {K.~S.}\ \bibnamefont
  {McCann}},\ }\href@noop {} {\bibfield  {journal} {\bibinfo  {journal}
  {Nature}\ }\textbf {\bibinfo {volume} {405}},\ \bibinfo {pages} {228}
  (\bibinfo {year} {2000})}\BibitemShut {NoStop}%
\bibitem [{\citenamefont {Ives}\ and\ \citenamefont {Carpenter}(2007)}]{ives}%
  \BibitemOpen
  \bibfield  {author} {\bibinfo {author} {\bibfnamefont {A.~R.}\ \bibnamefont
  {Ives}}\ and\ \bibinfo {author} {\bibfnamefont {S.~R.}\ \bibnamefont
  {Carpenter}},\ }\href {\doibase 10.1126/science.1133258} {\bibfield
  {journal} {\bibinfo  {journal} {Science}\ }\textbf {\bibinfo {volume}
  {317}},\ \bibinfo {pages} {58} (\bibinfo {year} {2007})}\BibitemShut
  {NoStop}%
\bibitem [{\citenamefont {Itoh}(1971)}]{itoh1971boltzmann}%
  \BibitemOpen
  \bibfield  {author} {\bibinfo {author} {\bibfnamefont {Y.}~\bibnamefont
  {Itoh}},\ }\href@noop {} {\bibfield  {journal} {\bibinfo  {journal}
  {Proceedings of the Japan Academy}\ }\textbf {\bibinfo {volume} {47}},\
  \bibinfo {pages} {854} (\bibinfo {year} {1971})}\BibitemShut {NoStop}%
\bibitem [{\citenamefont {May}\ and\ \citenamefont
  {Leonard}(1975)}]{may1975nonlinear}%
  \BibitemOpen
  \bibfield  {author} {\bibinfo {author} {\bibfnamefont {R.~M.}\ \bibnamefont
  {May}}\ and\ \bibinfo {author} {\bibfnamefont {W.~J.}\ \bibnamefont
  {Leonard}},\ }\href@noop {} {\bibfield  {journal} {\bibinfo  {journal} {SIAM
  journal on applied mathematics}\ }\textbf {\bibinfo {volume} {29}},\ \bibinfo
  {pages} {243} (\bibinfo {year} {1975})}\BibitemShut {NoStop}%
\bibitem [{\citenamefont {Hofbauer}\ and\ \citenamefont
  {Sigmund}(1998)}]{hofbauer1998evolutionary}%
  \BibitemOpen
  \bibfield  {author} {\bibinfo {author} {\bibfnamefont {J.}~\bibnamefont
  {Hofbauer}}\ and\ \bibinfo {author} {\bibfnamefont {K.}~\bibnamefont
  {Sigmund}},\ }\href@noop {} {\emph {\bibinfo {title} {Evolutionary games and
  population dynamics}}}\ (\bibinfo  {publisher} {Cambridge university press},\
  \bibinfo {year} {1998})\BibitemShut {NoStop}%
\bibitem [{\citenamefont {Reichenbach}\ \emph {et~al.}(2006)\citenamefont
  {Reichenbach}, \citenamefont {Mobilia},\ and\ \citenamefont
  {Frey}}]{reichenbach2006coexistence}%
  \BibitemOpen
  \bibfield  {author} {\bibinfo {author} {\bibfnamefont {T.}~\bibnamefont
  {Reichenbach}}, \bibinfo {author} {\bibfnamefont {M.}~\bibnamefont
  {Mobilia}}, \ and\ \bibinfo {author} {\bibfnamefont {E.}~\bibnamefont
  {Frey}},\ }\href@noop {} {\bibfield  {journal} {\bibinfo  {journal} {Physical
  Review E}\ }\textbf {\bibinfo {volume} {74}},\ \bibinfo {pages} {051907}
  (\bibinfo {year} {2006})}\BibitemShut {NoStop}%
\bibitem [{\citenamefont {Claussen}\ and\ \citenamefont
  {Traulsen}(2008)}]{claussen2008cyclic}%
  \BibitemOpen
  \bibfield  {author} {\bibinfo {author} {\bibfnamefont {J.~C.}\ \bibnamefont
  {Claussen}}\ and\ \bibinfo {author} {\bibfnamefont {A.}~\bibnamefont
  {Traulsen}},\ }\href@noop {} {\bibfield  {journal} {\bibinfo  {journal}
  {Physical review letters}\ }\textbf {\bibinfo {volume} {100}},\ \bibinfo
  {pages} {058104} (\bibinfo {year} {2008})}\BibitemShut {NoStop}%
\bibitem [{\citenamefont {Dayton}(1971)}]{dayton}%
  \BibitemOpen
  \bibfield  {author} {\bibinfo {author} {\bibfnamefont {P.}~\bibnamefont
  {Dayton}},\ }\href@noop {} {\bibfield  {journal} {\bibinfo  {journal} {Ecol.
  Monogr.}\ }\textbf {\bibinfo {volume} {41}},\ \bibinfo {pages} {351}
  (\bibinfo {year} {1971})}\BibitemShut {NoStop}%
\bibitem [{\citenamefont {Gilpin}(1975)}]{gilpin}%
  \BibitemOpen
  \bibfield  {author} {\bibinfo {author} {\bibfnamefont {M.~E.}\ \bibnamefont
  {Gilpin}},\ }\href {http://www.jstor.org/stable/2459636} {\bibfield
  {journal} {\bibinfo  {journal} {Am. Nat.}\ }\textbf {\bibinfo {volume}
  {109}},\ \bibinfo {pages} {51} (\bibinfo {year} {1975})}\BibitemShut
  {NoStop}%
\bibitem [{\citenamefont {Jackson}\ and\ \citenamefont
  {Buss}(1975)}]{jackson-1975}%
  \BibitemOpen
  \bibfield  {author} {\bibinfo {author} {\bibfnamefont {J.}~\bibnamefont
  {Jackson}}\ and\ \bibinfo {author} {\bibfnamefont {L.}~\bibnamefont {Buss}},\
  }\href@noop {} {\bibfield  {journal} {\bibinfo  {journal} {Proceedings of the
  National Academy of Sciences}\ }\textbf {\bibinfo {volume} {72}},\ \bibinfo
  {pages} {5160} (\bibinfo {year} {1975})}\BibitemShut {NoStop}%
\bibitem [{\citenamefont {Karlson}\ and\ \citenamefont {Buss}(1984)}]{karlson}%
  \BibitemOpen
  \bibfield  {author} {\bibinfo {author} {\bibfnamefont {R.}~\bibnamefont
  {Karlson}}\ and\ \bibinfo {author} {\bibfnamefont {L.}~\bibnamefont {Buss}},\
  }\href@noop {} {\bibfield  {journal} {\bibinfo  {journal} {Ecological
  Modeling}\ }\textbf {\bibinfo {volume} {23}},\ \bibinfo {pages} {243}
  (\bibinfo {year} {1984})}\BibitemShut {NoStop}%
\bibitem [{\citenamefont {Tainaka}(1988)}]{tainaka1988lattice}%
  \BibitemOpen
  \bibfield  {author} {\bibinfo {author} {\bibfnamefont {K.-i.}\ \bibnamefont
  {Tainaka}},\ }\href@noop {} {\bibfield  {journal} {\bibinfo  {journal}
  {Journal of the Physical Society of Japan}\ }\textbf {\bibinfo {volume}
  {57}},\ \bibinfo {pages} {2588} (\bibinfo {year} {1988})}\BibitemShut
  {NoStop}%
\bibitem [{\citenamefont {Boerlijst}\ and\ \citenamefont
  {Hogeweg}(1991)}]{boerlijst-1991}%
  \BibitemOpen
  \bibfield  {author} {\bibinfo {author} {\bibfnamefont {M.~C.}\ \bibnamefont
  {Boerlijst}}\ and\ \bibinfo {author} {\bibfnamefont {P.}~\bibnamefont
  {Hogeweg}},\ }\href@noop {} {\bibfield  {journal} {\bibinfo  {journal}
  {Physica D: Nonlinear Phenomena}\ }\textbf {\bibinfo {volume} {48}},\
  \bibinfo {pages} {17} (\bibinfo {year} {1991})}\BibitemShut {NoStop}%
\bibitem [{\citenamefont {Cronhjort}(1995)}]{cronhjort}%
  \BibitemOpen
  \bibfield  {author} {\bibinfo {author} {\bibfnamefont {M.~B.}\ \bibnamefont
  {Cronhjort}},\ }\href@noop {} {\bibfield  {journal} {\bibinfo  {journal}
  {Orig Life Evol Biosph}\ }\textbf {\bibinfo {volume} {25}},\ \bibinfo {pages}
  {227} (\bibinfo {year} {1995})}\BibitemShut {NoStop}%
\bibitem [{\citenamefont {Szabo}\ and\ \citenamefont
  {Czaran}(2001)}]{Szabo2001}%
  \BibitemOpen
  \bibfield  {author} {\bibinfo {author} {\bibfnamefont {G.}~\bibnamefont
  {Szabo}}\ and\ \bibinfo {author} {\bibfnamefont {T.}~\bibnamefont {Czaran}},\
  }\href@noop {} {\bibfield  {journal} {\bibinfo  {journal} {Phys. Rev. E}\
  }\textbf {\bibinfo {volume} {64}},\ \bibinfo {pages} {042902} (\bibinfo
  {year} {2001})}\BibitemShut {NoStop}%
\bibitem [{\citenamefont {Kerr}\ \emph
  {et~al.}(2002{\natexlab{a}})\citenamefont {Kerr}, \citenamefont {Riley},
  \citenamefont {Feldman},\ and\ \citenamefont {Bohannan}}]{kerr-2002}%
  \BibitemOpen
  \bibfield  {author} {\bibinfo {author} {\bibfnamefont {B.}~\bibnamefont
  {Kerr}}, \bibinfo {author} {\bibfnamefont {M.~A.}\ \bibnamefont {Riley}},
  \bibinfo {author} {\bibfnamefont {M.~W.}\ \bibnamefont {Feldman}}, \ and\
  \bibinfo {author} {\bibfnamefont {B.~J.}\ \bibnamefont {Bohannan}},\
  }\href@noop {} {\bibfield  {journal} {\bibinfo  {journal} {Nature}\ }\textbf
  {\bibinfo {volume} {418}},\ \bibinfo {pages} {171} (\bibinfo {year}
  {2002}{\natexlab{a}})}\BibitemShut {NoStop}%
\bibitem [{\citenamefont {Murrell}\ and\ \citenamefont
  {Law}(2003)}]{murrell-2003}%
  \BibitemOpen
  \bibfield  {author} {\bibinfo {author} {\bibfnamefont {D.~J.}\ \bibnamefont
  {Murrell}}\ and\ \bibinfo {author} {\bibfnamefont {R.}~\bibnamefont {Law}},\
  }\href@noop {} {\bibfield  {journal} {\bibinfo  {journal} {Ecology letters}\
  }\textbf {\bibinfo {volume} {6}},\ \bibinfo {pages} {48} (\bibinfo {year}
  {2003})}\BibitemShut {NoStop}%
\bibitem [{\citenamefont {Kirkup}\ and\ \citenamefont
  {Riley}(2004)}]{kirkup2004antibiotic}%
  \BibitemOpen
  \bibfield  {author} {\bibinfo {author} {\bibfnamefont {B.~C.}\ \bibnamefont
  {Kirkup}}\ and\ \bibinfo {author} {\bibfnamefont {M.~A.}\ \bibnamefont
  {Riley}},\ }\href@noop {} {\bibfield  {journal} {\bibinfo  {journal}
  {Nature}\ }\textbf {\bibinfo {volume} {428}},\ \bibinfo {pages} {412}
  (\bibinfo {year} {2004})}\BibitemShut {NoStop}%
\bibitem [{\citenamefont {Laird}\ and\ \citenamefont
  {Schamp}(2006)}]{Laird2006}%
  \BibitemOpen
  \bibfield  {author} {\bibinfo {author} {\bibfnamefont {R.}~\bibnamefont
  {Laird}}\ and\ \bibinfo {author} {\bibfnamefont {B.}~\bibnamefont {Schamp}},\
  }\href@noop {} {\bibfield  {journal} {\bibinfo  {journal} {Am. Nat.}\
  }\textbf {\bibinfo {volume} {168}},\ \bibinfo {pages} {182} (\bibinfo {year}
  {2006})}\BibitemShut {NoStop}%
\bibitem [{\citenamefont {Kerr}\ \emph {et~al.}(2006)\citenamefont {Kerr},
  \citenamefont {Neuhauser}, \citenamefont {Bohannan},\ and\ \citenamefont
  {Dean}}]{kerr}%
  \BibitemOpen
  \bibfield  {author} {\bibinfo {author} {\bibfnamefont {B.}~\bibnamefont
  {Kerr}}, \bibinfo {author} {\bibfnamefont {C.}~\bibnamefont {Neuhauser}},
  \bibinfo {author} {\bibfnamefont {B.~J.~M.}\ \bibnamefont {Bohannan}}, \ and\
  \bibinfo {author} {\bibfnamefont {A.~M.}\ \bibnamefont {Dean}},\ }\href
  {\doibase 10.1038/nature04864} {\bibfield  {journal} {\bibinfo  {journal}
  {Nature}\ }\textbf {\bibinfo {volume} {442}},\ \bibinfo {pages} {75}
  (\bibinfo {year} {2006})}\BibitemShut {NoStop}%
\bibitem [{\citenamefont {Reichenbach}\ and\ \citenamefont
  {Frey}(2008)}]{reichenbach2008instability}%
  \BibitemOpen
  \bibfield  {author} {\bibinfo {author} {\bibfnamefont {T.}~\bibnamefont
  {Reichenbach}}\ and\ \bibinfo {author} {\bibfnamefont {E.}~\bibnamefont
  {Frey}},\ }\href@noop {} {\bibfield  {journal} {\bibinfo  {journal} {Physical
  review letters}\ }\textbf {\bibinfo {volume} {101}},\ \bibinfo {pages}
  {058102} (\bibinfo {year} {2008})}\BibitemShut {NoStop}%
\bibitem [{\citenamefont {Rulands}\ \emph {et~al.}(2013)\citenamefont
  {Rulands}, \citenamefont {Zielinski},\ and\ \citenamefont
  {Frey}}]{rulands2013global}%
  \BibitemOpen
  \bibfield  {author} {\bibinfo {author} {\bibfnamefont {S.}~\bibnamefont
  {Rulands}}, \bibinfo {author} {\bibfnamefont {A.}~\bibnamefont {Zielinski}},
  \ and\ \bibinfo {author} {\bibfnamefont {E.}~\bibnamefont {Frey}},\
  }\href@noop {} {\bibfield  {journal} {\bibinfo  {journal} {Physical Review
  E}\ }\textbf {\bibinfo {volume} {87}},\ \bibinfo {pages} {052710} (\bibinfo
  {year} {2013})}\BibitemShut {NoStop}%
\bibitem [{\citenamefont {Szolnoki}\ and\ \citenamefont
  {Perc}(2016)}]{szolnoki2016zealots}%
  \BibitemOpen
  \bibfield  {author} {\bibinfo {author} {\bibfnamefont {A.}~\bibnamefont
  {Szolnoki}}\ and\ \bibinfo {author} {\bibfnamefont {M.}~\bibnamefont
  {Perc}},\ }\href@noop {} {\bibfield  {journal} {\bibinfo  {journal} {Physical
  Review E}\ }\textbf {\bibinfo {volume} {93}},\ \bibinfo {pages} {062307}
  (\bibinfo {year} {2016})}\BibitemShut {NoStop}%
\bibitem [{\citenamefont {Mathiesen}\ \emph {et~al.}(2011)\citenamefont
  {Mathiesen}, \citenamefont {Mitarai}, \citenamefont {Sneppen},\ and\
  \citenamefont {Trusina}}]{mathiesen2011ecosystems}%
  \BibitemOpen
  \bibfield  {author} {\bibinfo {author} {\bibfnamefont {J.}~\bibnamefont
  {Mathiesen}}, \bibinfo {author} {\bibfnamefont {N.}~\bibnamefont {Mitarai}},
  \bibinfo {author} {\bibfnamefont {K.}~\bibnamefont {Sneppen}}, \ and\
  \bibinfo {author} {\bibfnamefont {A.}~\bibnamefont {Trusina}},\ }\href@noop
  {} {\bibfield  {journal} {\bibinfo  {journal} {Physical review letters}\
  }\textbf {\bibinfo {volume} {107}},\ \bibinfo {pages} {188101} (\bibinfo
  {year} {2011})}\BibitemShut {NoStop}%
\bibitem [{\citenamefont {Harris}(1996)}]{harris1996competitive}%
  \BibitemOpen
  \bibfield  {author} {\bibinfo {author} {\bibfnamefont {P.~M.}\ \bibnamefont
  {Harris}},\ }\href@noop {} {\bibfield  {journal} {\bibinfo  {journal}
  {Oecologia}\ }\textbf {\bibinfo {volume} {108}},\ \bibinfo {pages} {663}
  (\bibinfo {year} {1996})}\BibitemShut {NoStop}%
\bibitem [{\citenamefont {Nash}(2008)}]{N08}%
  \BibitemOpen
  \bibfield  {author} {\bibinfo {author} {\bibfnamefont {T.~H.}\ \bibnamefont
  {Nash}},\ }\href@noop {} {\emph {\bibinfo {title} {Lichen Biology}}}\
  (\bibinfo  {publisher} {Cambridge University Press},\ \bibinfo {year}
  {2008})\BibitemShut {NoStop}%
\bibitem [{\citenamefont {Grube}\ \emph {et~al.}(2009)\citenamefont {Grube},
  \citenamefont {Cardinale}, \citenamefont {de~Castro}, \citenamefont
  {Müller},\ and\ \citenamefont {Berg}}]{G09}%
  \BibitemOpen
  \bibfield  {author} {\bibinfo {author} {\bibfnamefont {M.}~\bibnamefont
  {Grube}}, \bibinfo {author} {\bibfnamefont {M.}~\bibnamefont {Cardinale}},
  \bibinfo {author} {\bibfnamefont {J.~V.}\ \bibnamefont {de~Castro}}, \bibinfo
  {author} {\bibfnamefont {H.}~\bibnamefont {Müller}}, \ and\ \bibinfo
  {author} {\bibfnamefont {G.}~\bibnamefont {Berg}},\ }\href {\doibase
  10.1038/ismej.2009.63} {\bibfield  {journal} {\bibinfo  {journal} {ISME J}\
  }\textbf {\bibinfo {volume} {3}},\ \bibinfo {pages} {1105} (\bibinfo {year}
  {2009})}\BibitemShut {NoStop}%
\bibitem [{\citenamefont {Jettestuen}\ \emph {et~al.}(2010)\citenamefont
  {Jettestuen}, \citenamefont {Nermoen}, \citenamefont {Hestmark},
  \citenamefont {Timdal},\ and\ \citenamefont {Mathiesen}}]{JNHTM10}%
  \BibitemOpen
  \bibfield  {author} {\bibinfo {author} {\bibfnamefont {E.}~\bibnamefont
  {Jettestuen}}, \bibinfo {author} {\bibfnamefont {A.}~\bibnamefont {Nermoen}},
  \bibinfo {author} {\bibfnamefont {G.}~\bibnamefont {Hestmark}}, \bibinfo
  {author} {\bibfnamefont {E.}~\bibnamefont {Timdal}}, \ and\ \bibinfo {author}
  {\bibfnamefont {J.}~\bibnamefont {Mathiesen}},\ }\href {\doibase
  10.1371/journal.pone.0012820} {\bibfield  {journal} {\bibinfo  {journal}
  {PLoS One}\ }\textbf {\bibinfo {volume} {5}},\ \bibinfo {pages} {e12820}
  (\bibinfo {year} {2010})}\BibitemShut {NoStop}%
\bibitem [{\citenamefont {Aerts}\ and\ \citenamefont {Soest}(1997)}]{AS1997}%
  \BibitemOpen
  \bibfield  {author} {\bibinfo {author} {\bibfnamefont {L.}~\bibnamefont
  {Aerts}}\ and\ \bibinfo {author} {\bibfnamefont {R.}~\bibnamefont {Soest}},\
  }\href@noop {} {\bibfield  {journal} {\bibinfo  {journal} {Mar. Ecol. Prog.
  Ser.}\ }\textbf {\bibinfo {volume} {148}},\ \bibinfo {pages} {125} (\bibinfo
  {year} {1997})}\BibitemShut {NoStop}%
\bibitem [{\citenamefont {Buss}\ and\ \citenamefont {Jackson}(1979)}]{buss}%
  \BibitemOpen
  \bibfield  {author} {\bibinfo {author} {\bibfnamefont {L.}~\bibnamefont
  {Buss}}\ and\ \bibinfo {author} {\bibfnamefont {J.}~\bibnamefont {Jackson}},\
  }\href@noop {} {\bibfield  {journal} {\bibinfo  {journal} {Am. Nat.}\
  }\textbf {\bibinfo {volume} {113}},\ \bibinfo {pages} {223} (\bibinfo {year}
  {1979})}\BibitemShut {NoStop}%
\bibitem [{\citenamefont {Mitarai}\ \emph {et~al.}(2012)\citenamefont
  {Mitarai}, \citenamefont {Mathiesen},\ and\ \citenamefont
  {Sneppen}}]{mitarai2012emergence}%
  \BibitemOpen
  \bibfield  {author} {\bibinfo {author} {\bibfnamefont {N.}~\bibnamefont
  {Mitarai}}, \bibinfo {author} {\bibfnamefont {J.}~\bibnamefont {Mathiesen}},
  \ and\ \bibinfo {author} {\bibfnamefont {K.}~\bibnamefont {Sneppen}},\
  }\href@noop {} {\bibfield  {journal} {\bibinfo  {journal} {Physical Review
  E}\ }\textbf {\bibinfo {volume} {86}},\ \bibinfo {pages} {011929} (\bibinfo
  {year} {2012})}\BibitemShut {NoStop}%
\bibitem [{\citenamefont {Botta}\ and\ \citenamefont
  {Mitarai}(2014)}]{botta2014disturbance}%
  \BibitemOpen
  \bibfield  {author} {\bibinfo {author} {\bibfnamefont {F.}~\bibnamefont
  {Botta}}\ and\ \bibinfo {author} {\bibfnamefont {N.}~\bibnamefont
  {Mitarai}},\ }\href@noop {} {\bibfield  {journal} {\bibinfo  {journal}
  {Physical Review E}\ }\textbf {\bibinfo {volume} {89}},\ \bibinfo {pages}
  {022704} (\bibinfo {year} {2014})}\BibitemShut {NoStop}%
\bibitem [{\citenamefont {Mitarai}\ \emph {et~al.}(2014)\citenamefont
  {Mitarai}, \citenamefont {Heinsalu},\ and\ \citenamefont
  {Sneppen}}]{mitarai2014speciation}%
  \BibitemOpen
  \bibfield  {author} {\bibinfo {author} {\bibfnamefont {N.}~\bibnamefont
  {Mitarai}}, \bibinfo {author} {\bibfnamefont {E.}~\bibnamefont {Heinsalu}}, \
  and\ \bibinfo {author} {\bibfnamefont {K.}~\bibnamefont {Sneppen}},\
  }\href@noop {} {\bibfield  {journal} {\bibinfo  {journal} {PloS one}\
  }\textbf {\bibinfo {volume} {9}},\ \bibinfo {pages} {e96665} (\bibinfo {year}
  {2014})}\BibitemShut {NoStop}%
\bibitem [{\citenamefont {Kay}\ and\ \citenamefont
  {Keough}(1981)}]{KayandKeough1981}%
  \BibitemOpen
  \bibfield  {author} {\bibinfo {author} {\bibfnamefont {A.~M.}\ \bibnamefont
  {Kay}}\ and\ \bibinfo {author} {\bibfnamefont {M.~J.}\ \bibnamefont
  {Keough}},\ }\href@noop {} {\bibfield  {journal} {\bibinfo  {journal}
  {Oecologia}\ }\textbf {\bibinfo {volume} {48}},\ \bibinfo {pages} {73}
  (\bibinfo {year} {1981})}\BibitemShut {NoStop}%
\bibitem [{\citenamefont {Cz{\'a}r{\'a}n}\ \emph {et~al.}(2002)\citenamefont
  {Cz{\'a}r{\'a}n}, \citenamefont {Hoekstra},\ and\ \citenamefont
  {Pagie}}]{czaran2002chemical}%
  \BibitemOpen
  \bibfield  {author} {\bibinfo {author} {\bibfnamefont {T.~L.}\ \bibnamefont
  {Cz{\'a}r{\'a}n}}, \bibinfo {author} {\bibfnamefont {R.~F.}\ \bibnamefont
  {Hoekstra}}, \ and\ \bibinfo {author} {\bibfnamefont {L.}~\bibnamefont
  {Pagie}},\ }\href@noop {} {\bibfield  {journal} {\bibinfo  {journal}
  {Proceedings of the National Academy of Sciences}\ }\textbf {\bibinfo
  {volume} {99}},\ \bibinfo {pages} {786} (\bibinfo {year} {2002})}\BibitemShut
  {NoStop}%
\bibitem [{\citenamefont {Kerr}\ \emph
  {et~al.}(2002{\natexlab{b}})\citenamefont {Kerr}, \citenamefont {Riley},
  \citenamefont {Feldman},\ and\ \citenamefont {Bohannan}}]{kerr2002}%
  \BibitemOpen
  \bibfield  {author} {\bibinfo {author} {\bibfnamefont {B.}~\bibnamefont
  {Kerr}}, \bibinfo {author} {\bibfnamefont {M.~A.}\ \bibnamefont {Riley}},
  \bibinfo {author} {\bibfnamefont {M.~W.}\ \bibnamefont {Feldman}}, \ and\
  \bibinfo {author} {\bibfnamefont {B.~J.~M.}\ \bibnamefont {Bohannan}},\
  }\href {\doibase 10.1038/nature00823} {\bibfield  {journal} {\bibinfo
  {journal} {Nature}\ }\textbf {\bibinfo {volume} {418}},\ \bibinfo {pages}
  {171} (\bibinfo {year} {2002}{\natexlab{b}})}\BibitemShut {NoStop}%
\end{thebibliography}%
\bibliographystyle{apsrev4-1}

\end{document}